# Soft X-ray absorption spectroscopy and magnetic circular dichroism under pulsed high magnetic field of Ni-Co-Mn-In metamagnetic shape memory alloy

R.Y. Umetsu [1]*, H. Yasumura [1], Y. Narumi [2], Y. Kotani [3], T. Nakamura [3,4,5], H. Nojiri [1] and R. Kainuma [6]

1 Institute for Materials Research, Tohoku University, Sendai 980-8577, Japan
2 Center for Advanced High Magnetic Field Science, Graduate School of Science, Osaka University, Toyonaka 560-0043, Japan
3 Center for Synchrotron Radiation Research, Japan Synchrotron Radiation Research Institute, SPring-8, Hyogo 679-5198, Japan
4 Institute of Multidisciplinary Research for Advanced Materials, Tohoku University, Sendai 980-8577, Japan
5 International Center for Synchrotron Radiation Innovation Smart, Tohoku University, Sendai 980-8577, Japan
6 Department of Materials Science, Graduate School of Engineering, Tohoku University, Sendai 980-8579, Japan

*Corresponding author, *E-mail address*: rieume@imr.tohoku.ac.jp (R.Y. Umetsu)

**Abstract**

In this study, X-ray absorption spectroscopy (XAS) experiments for $Ni_{45}Co_5Mn_{36.7}In_{13.3}$ metamagnetic shape memory alloy were performed under high magnetic fields up to 12 T using a pulsed magnet. Field–induced reverse transformation to austenite phase caused considerable changes in the magnetic circular dichroism (MCD) signals and the magnetic moments of the ferromagnetic coupling between Mn, Ni, and Co were determined. The spin magnetic moment, $M_{spin}$, and orbital magnetic moment, $M_{orb}$, of Mn atom in the induced austenite ferromagnetic phase, estimated based on the magneto-optical sum rule, were 3.2 and 0.13 $\mu_B$, respectively, resulting in an $M_{orb}$ / $M_{spin}$ ratio of 0.04. In the element-specific magnetization curves recorded at 150 K, metamagnetic behavior associated with the field–induced reverse transformation is clearly observed and reverse transformation finishing magnetic field and martensitic transformation starting magnetic field are detected. There was almost no difference in the magnetically averaged XAS spectrum for Mn-$L_{2,3}$ edges between in the martensite and in the magnetic field-induced austenite phases, however, it was visible for Ni, indicating that Ni 3$d$-electrons mainly contribute to martensitic transformation.

# I. Introduction

Prototype of ferromagnetic shape memory material is $Ni_2MnGa$ Heusler alloy. The occurrence of martensitic transformation was first reported in a study on neutron diffraction by Webster *et al*. [1]. They found that the $L2_1$-type ordered cubic structure in the austenite phase transforms into a complex tetragonal structure with lower crystal symmetry in the martensite phase. A huge magnetic strain in the martensite phase was observed in the single crystal of $Ni_2MnGa$ [2,3]. Consequently, a large number of studies have been conducted owing to the favorable efficiency of the material in actuators and sensors so that numerous application [4-6]. The unique behavior of martensitic phase transformation was first reported in off-stoichiometric Heusler Ni-Mn-$Z$ ($Z$ = In, Sn, and Sb) alloys by Sutou *et al*. [7]. These alloys have also attracted widespread interest as high-performance multiferroic materials. Besides the structural changes due to martensitic transformation, magnetic properties drastically change simultaneously with the transformation from the ferromagnetic austenite phase to the martensite phase with weak magnetization. Thus, Zeeman energy caused by applying a magnetic field stabilizes the austenite phase and the transformation is also controlled by the magnetic field, whereas the general martensitic transformation is controlled by temperature and/or stress. Substituting Co with Ni enhances magnetization in the austenite phase and decreases it in the martensite phase, resulting in a larger change in magnetization between the two phases. According to previous studies, the martensitic transformation in optimized $Ni_{45}Co_5Mn_{36.7}In_{13.3}$ decreases at the rate of about 4.3 Kelvin/Tesla, and metamagnetic behavior is caused by field-induced reverse transformation below the martensitic transformation temperature [8]. Apart from the metamagnetic behavior, the alloys show many interesting physical properties, including magnetic shape memory effect [8,9], inverse magnetocaloric effect [10–12], giant magnetoresistance effect [13,14], and giant magnetothermal conductivity [15].

X-ray magnetic circular dichroism (MCD), obtained by X-ray absorption spectroscopy (XAS), is one of the magneto-optical effects and provides element-specific information, such as magnetic moments, where element selectivity follows the selection rule of core-electron excitation. In particular, MCD in the soft X-ray region possesses the inherent advantage of probing the $3d$ bands in transition metals. Several studies on MCD have been reported for $Ni_2MnGa$ [16,17], and related materials, such as off-stoichiometric composition [18] and Cu- or Co-doped alloys [19,20]. However, only two studies have been reported on the metamagnetic behavior by XAS-MCD of Ni-Mn-In bulk system [21,22] and one on that of Ni-Co-Mn-In film [23]. All the studies were restricted to MCD observation of the martensitic transformation driven by temperature changes, because a high magnetic field is required for the direct observation of field-induced reverse transformation.

So far, the highest static magnetic field for MCD experiments is 17 T using a superconducting solenoid magnet, installed at the ID12 beam line in ESRF [24]. As a more specialized technique, MCD of up to 40 T in the hard X-ray region has been established with a pulsed magnet using a portable capacitor bank [25,26]. A simple transmission method in the atmosphere can be used for the long penetration depth in the hard X-ray region. On the other hand, an ultra-high vacuum (UHV) system for soft X-ray absorption measurements was realized using a UHV chamber with a pulsed magnet coil because it is required to investigate the $3d$-electron bands in transition metal alloys and compounds [27,28]. Using a pulsed high magnetic field soft X-ray MCD apparatus equipped with a newly developed bipolar capacitor bank, field-induced spin reorientation was successfully observed in ferrimagnetic spinel $MnCr_2S_4$ [29]. We present here the results of XAS-MCD experiments for the field-induced reverse martensitic transformation in the Ni-Co-Mn-In Heusler alloy. We compare and discuss the difference of the XAS spectra between in the martensite and austenite phases directly without elevating the temperature.

**II. Experimental Procedure**

$Ni_{45}Co_5Mn_{36.7}In_{13.3}$ alloy was fabricated by induction melting in an Ar atmosphere. The specimen was sealed in a quartz capsule and annealed at 1173 K for 1 day and then quenched in water. To lower the transformation temperature, the specimen was further annealed at 673 K for 3 h [30]. The magnetic measurements were conducted using a superconducting quantum interference device (SQUID) magnetometer up to 5 T magnetic fields and by the extraction method using a superconducting magnet up to 18 T. The measurements were performed in the High Field Laboratory for Superconducting Materials, Institute for Materials Research, Tohoku University.

XAS-MCD measurements were conducted using the total electron yield (TEY) method at soft X-ray beamline BL25SU, SPring-8, Japan. At BL25SU, circularly polarized soft X-ray was generated using a twin helical undulator [31]. A non-destructive pulsed magnet capable of generating 30 T was mounted outside the UHV chamber and directly cooled by liquid nitrogen. The duration of the pulsed magnetic fields was about 50 ms with the asymmetric pulse shape. The TEY signal and the induction voltage from the magnetic field were simultaneously recorded at a frequency of 1 MHz. The details of the experimental setup are available in Ref. [27,28]. The specimen was cut into strips of $1 \times 1$ mm$^2$ cross-section and 5 mm length. The strip was fixed on the sapphire sample stage with the electrically conductive epoxy adhesive and cleaved inside the UHV chamber to obtain a fresh surface for the XAS-MCD experiments.

**III. Results and Discussion**

**A. Magnetic Measurements**

Figures 1 (a) and (b) show thermomagnetization (*M-T*) curves of $Ni_{45}Co_5Mn_{36.7}In_{13.3}$ measured at 0.05, 1, and 3 T, and 5 and 9 T, respectively. The *M-T* curves were recorded under field-heating and field-cooling processes at each magnetic field. Thermal hysteresis and large change in the magnetization due to martensitic transformation are observed in all *M-T* curves. In the figures, $T_{Af}$ and $T_{Ms}$ denote reverse transformation finishing temperature to austenite phase and the martensitic transformation starting temperature, respectively. $T_{Af}$ and $T_{Ms}$ at 0.05 T magnetic field are 282 and 249 K, respectively, which are comparable to those in our previous report [32]. The martensitic transformation temperatures decrease with increasing the applied magnetic field because the ferromagnetic austenite phase is stabilized by the Zeeman energy. $T_{Af}$ at 5 T magnetic field is 247 K, thus, the transformation temperature decreases at the rate of 7 K/T. Assuming the change in magnetization between the austenite and martensite phases to be about 120 A·m$^2$/kg, as shown in Fig. 1 (a), the entropy change, $\Delta S$, due to the martensitic transformation is estimated to be 17.1 J/kg-K from the Clausius-Clapeyron equation, $\Delta S = (dH/dT) \cdot \Delta M \approx (\Delta H/\Delta T) \cdot \Delta M$. Magnetization under the field cooling process at 5 T and 9 T remained relatively high even at low temperatures. This is thermally arrested behavior, where the martensitic transformation does not occur or is not well complete [33,34].

Figures 2 (a) and (b) are magnetization (*M-H*) curves obtained at 100, 150, and 200 K and 7, 50, and 100 K, respectively. $H_{Af}$ and $H_{Ms}$ denote the field-induced transformation finishing magnetic field to austenite phase and martensitic phase transformation starting magnetic field, respectively, and they are defined as the intersection points of the extrapolated lines in the *M-H* curves. As shown in Fig. 2 (a), the measuring temperatures are comparatively close to the martensitic transformation temperature; thus, the critical magnetic fields of $H_{Af}$ and $H_{Ms}$ increase with decreasing temperature. However, in Fig. 2 (b), $H_{Af}$ increases but $H_{Ms}$ decreases with decreasing temperature. This shows that the magnetic hysteresis expands with decreasing temperature. In the *M-H* curve measured at 7 K, $H_{Af}$ is clearly detected, however, it was hard to decide the $H_{Ms}$ because of the large hysteresis.

Temperature dependence of the critical magnetic fields of $H_{Af}$ and $H_{Ms}$ determined from the *M-H* curves in Fig. 2 is plotted in Fig. 3. Here, $H_0$ is defined as $H_0 = (H_{Af} + H_{Ms})/2$. The martensitic transformation temperatures, $T_{Af}$, $T_{Ms}$, and $T_0 = (T_{Af} + T_{Ms})/2$ obtained from the *M-T* curves in Fig. 1 are also plotted in the same figure. The critical magnetic fields and temperatures show a horn shape behavior. The whole shapes are consistent with those of our previous report [33], although the final annealing temperature was slightly different. Here, $H_0$ and $T_0$ were constant below 130 K. The constant value of $H_0$ means that its derivative with respect to temperature ($dH_0/dT$) becomes to zero;

that is, the entropy change, $\varDelta S$, tends to zero, considering the Clausius-Clapeyron equation. Furthermore, XAS-MCD was investigated under magnetic fields of up to 12 T combine with a pulsed magnetic.

**B. X-ray absorption spectroscopy and magnetic circular dichroism under pulsed magnetic field**

Figures 4 (a) and (b) indicate the XAS and MCD spectra at Mn-$L_{2,3}$ edges for $Ni_{45}Co_5Mn_{36.7}In_{13.3}$ under the magnetic field up to 12 T with a pulsed magnet. Figure 4(a) shows the averaged XAS spectra, $XAS_{Av.} = (\mu^+ + \mu^-)/2$, under 1-3 and 10-12 T magnetic fields. The spectra were obtained at parallel and antiparallel configuration to the magnetization direction, $\mu^+$ and $\mu^-$, respectively. The $XAS_{Av.}$ spectra were accumulated from 1 to 3 T (10 to 12 T) to increase the statistics during a series of the magnetic field sweeping experiments. The measurements were performed at 150 K, which is enough lower than the martensitic transformation temperature and higher than the thermally arrested temperature of 130 K. From the $M$-$H$ curve at 150 K in Fig. 2 (a), the spectra at 1-3 T and 10-12 T correspond to that at the martensite and austenite phases, respectively. This is because a magnetic field of 10 T is needed to complete the field-induced reverse transformation from the martensite phase to the austenite phase. The $XAS_{Av}$ spectra for Mn-$L_{2,3}$ edges show multiplet structures, for example, shoulders in lower and higher energy sides in the $L_3$ edge main peak, a satellite around the photon energy of 644 eV, and doublet peaks in the $L_2$ edge, as indicated by solid bars. These structures have been discussed in terms of localized Mn magnetic moment due to Mn oxidization. However, they are currently considered as the intrinsic features of the localization of the Mn-$3d$ electrons. Similar behaviors have been reported in many Heusler alloys, such as $Co_2(Mn,Fe)Si$ [18], $Rh_2MnGe$ [35], $Ni_2MnGa$ [23, 36], and $Ni_2Mn_{1.4}In_{0.6}$ [21], and they are well reproduced by the calculated spectra. The inset of Fig. 4 (a) shows the expanded scale of the peak top area in the Mn-$L_3$ edge. The averaged spectra of $\mu^+$ and $\mu^-$ indicate the XAS spectra without magnetic contribution; that is, the difference in the $XAS_{Av}$ spectra between the before and after the transformation gives intrinsic information on the change in the electronic state related to the martensitic transformation. However, the spectra seem to be completely over layered, showing almost no change in the $XAS_{Av}$ spectra for Mn-$L_3$ edge. This is discussed in detail in a later section and compared with other elements.

Figure 4 (b) show the XAS spectra for $\mu^+$ and $\mu^-$, and the MCD (= $\mu^+$ - $\mu^-$) for 1-3, 4-6, 7-9, and 10-12 T. The data were accumulated for each range of magnetic field to include the statistics. A large difference between $\mu^+$ and $\mu^-$ is observed, indicating that the Mn-3$d$ electrons contribute to the magnetism of the alloy. In addition, the MCD intensities gradually increase with an increase in magnetic field, indicating the field-induced austenite phase exhibits ferromagnetism, whereas the martensite phase has weak-magnetism, thereby resulting in large MCD signals. This behavior corresponds to the metamagnetic transition observed in the *M-H* curve in Fig. 2 (a). Besides the Mn-$L_3$ edge main peak, the intensities of the MCD signals at the satellite positions observed around 644 eV and the doublet in the Mn-$L_2$ edge also increase gradually. This also indicates that multiple structures, showing shoulders and satellite, are associated with the magnetic character of Mn.

The $XAS_{Av}$ spectra for Ni-$L_{2,3}$ and Co-$L_{2,3}$ edges were collected and analyzed in the same way as the Mn-$L_{2,3}$ edges, as shown in Fig. 5 (a) and Fig. 6 (a), respectively. As shown in the insets, there is a clear difference in $XAS_{Av}$ in before and after the transformation. This is attributed to a certain change in the electronic state due to martensitic transformation, in contrast to the Mn-$L_3$ edge. In general, the decrease in the intensity of absorption corresponds to the decrease in the number of $d$ holes. In other words, the number of $d$ electrons in the ferromagnetic austenite phase is greater than that in the martensite phase. Furthermore, arrows denoted by A – D indicate the photon energies, which are pointed out to show certain changes in the density of states before and after the martensitic phase transformation, as suggested by Klaer *et al*. [23]. This is discussed in detail in a later section. Figures 5 (b) and 6 (b) indicate the XAS spectra for $\mu^+$ and $\mu^-$, and MCD for the Ni and Co-$L_{2,3}$ edges, respectively. The MCD signal increases with an increase in the magnetic field, and this trend is consistent with the *M-H* curves. Notably, MCD for the $L_3$ edge for all the elements is negative, the same as in Figs. 4-6 (b). This indicates that the magnetic moments of Mn, Ni, and Co are coupled ferromagnetically with each other. The composition of the specimen is off-stoichiometry of $Ni_{45}Co_5Mn_{36.7}In_{13.3}$; thus, it is speculated that 5 at.% of Co distributes in the Ni site with the 8*c* Wyckoff position and the excess Mn is located in the In site with 4*c* position. The Mn atoms have two different atomic sites: 4*c* and the ordinary Mn site with 4*b* position. It is inferred that excess Mn at the In site should have the same magnetic moment as that in the ordinary site in 4*b*, and the coupling between Mn at different sites should be ferromagnetic based on a systematic study on the concentration dependence of the magnetization of Ni-Mn-In alloys [37]. From the sum-rule analysis [38,39], the spin magnetic moments of Mn, Ni, and Co at 12 T are 3.19, 0.47, and 1.25 $\mu_B$/atom and the orbital magnetic moments are 0.08, 0.06, and 0.03 $\mu_B$, respectively, resulting in a total magnetic moment of 6.0 $\mu_B$/f.u. The saturation magnetization at 150 K is about 6.3 $\mu_B$/f.u. converted from 140 $Am^2$/kg, hence, they are in good agreement. The numerical values of the magnetic moments are listed in Table. The 3*d* electron numbers, $n_d$, for Mn, Ni, and Co are assumed to be 5.17, 8.50 and 7.51 respectively, and a correction factor of 1.5 is adopted for Mn [40,41] because of the mixing state of the $2p \rightarrow 3d$ excitation. $M_{orb}$ / $M_{spin}$ in the martensite phase with a lower symmetric structure tends to be larger than that in the austenite phase with a cubic structure in each element. For a further detailed discussion on the magnitude, data with higher accuracy are required.

Figure 7 (a) shows the element-specific hysteresis in Mn, Ni, and Co elements obtained at 150 K. The absolute values correspond to the magnitudes of the MCD intensities at the peak tops in each $L_3$ edge, where the photon energies are 639.6, 825.6, and 778.1 eV for Mn, Ni, and Co, respectively. For Ni and Co, the measurements were performed repeatedly and the intensities were accumulated because of the low accuracy due to the small MCD signals. For Mn and Ni, hysteresis loops associated with the field-induced reverse transformation are observed. Figure 7 (b) shows the *M-H* curve measured at 150 K as a reference, which is extracted from Fig. 2 (a). Magnetization in the martensite phase (under $H_{As}$) tends to increase with the increase in the applying magnetic field. The element-specific hysteresis curve of Mn also indicates the corresponding change. Comparing the figures, $H_{As}$ and $H_{Af}$ are slightly shifted to higher magnetic fields, whereas $H_{Ms}$ and $H_{Mf}$ in Fig. 7 (a) are lower than that in Fig. 7 (b). That is, the transformation hysteresis is expanded in the element-specific hysteresis than in the magnetic experiments. The reason for the difference has not been clarified; however, it is possibly due to surface effects. XAS results are generally sensitive to surface conditions, and the martensitic transformation temperature and/or transformation magnetic fields are affected by interfacial and surface energies [42,43]. Another possibility is the difference of the sample shape. A small grain was used in the magnetization measurements. On the other hand, polycrystalline specimen was used in the XAS experiments. Even if the information of the XAS spectra come from the single crystalline grain, the strain constraint from surrounding grains will have some effect on the transformation behaviour [44].

Figures 8 (a)-(c) show the differences in the averaged XAS spectra, $\Delta\mathrm{XAS}_{Av} = \mathrm{XAS}_{Av}(10\text{-}12\mathrm{T}) - \mathrm{XAS}_{Av}(1\text{-}3\mathrm{T})$ for Mn, Ni, and Co, as obtained from Fig. 4 (a), 5 (a), and 6 (a), respectively. Before the subtraction, the intensities of the peak top of the $\mathrm{XAS}_{AV}$ spectra at 1-3 T were normalized as unity. The system exhibits the martensite and austenite phases in 1-3 and 10-12 T, respectively, and $\Delta\mathrm{XAS}_{Av}$ corresponds to the difference in the electronic structure between the two phases. The cubic austenite phase transforms to a lower symmetry crystal structure, such as distorted and/or modulated long range structures, and the electronic structure could be changed by the martensitic transformation. In this study, the measurement temperature was fixed at 150 K, and the transformation was induced by applying a magnetic field. Therefore, the intrinsic difference in the $\mathrm{XAS}_{Av}$ spectra induced by the structural change could be observed without the temperature dependence of the spectra. As mentioned in the inset of Fig. 4(a), $\Delta\mathrm{XAS}_{Av}$ is invisible in Mn-$L_{2,3}$ edges. The solid bars are located at the same photon energies in Fig. 4 (a), indicating a characteristic feature of localized Mn moment in the $\mathrm{XAS}_{Av}$ spectra. However, at these photon energies, no noticeable feature is observed. From theoretical calculations on $Ni_2MnGa$, it was reported that there is no significant change in the partial density of states of Mn around the Fermi level in the martensite and austenite phases, whereas the tetragonal distortion changes the Ni-3$d$ states [45]. In fact, $\Delta XAS_{Av}$ in Ni is significant compared to that in Mn. The intensity of the peak top decreases due to the reverse phase transformation from martensite to the austenite phase, which corresponds to a decrease in the number of $d$ holes around the Fermi level, in other words, an increase in the number of $d$ electrons. Theoretical calculations have showed that DOS at the Fermi level in the austenite phase is larger than that in the martensite phase for Ni-Mn-In and Ni-Mn-Sn alloys [46,47]. It was also reported, from low-temperature specific heat measurements, that the electronic specific heat coefficient in the austenite phase is larger than that in the martensite phase for Ni-Co-Mn-In alloys [48], which is consistent with the present results.

In Fig. 8 (b), the arrows designated as A, B, C, and D indicate the same photon energies as Fig. 5 (a), which are also reported by Klaer *et al*.[23]. The double arrow indicates the photon energy corresponding to a peak top in the XAS. A is located in the pre-edge region and B and C at 1.5 and 6 eV above the $L_3$ edge peak position, respectively. Theoretical calculations of the DOS of Ni-based shape memory alloys revealed that Ni-$e_g$ orbitals are degenerated in the cubic phase but split in the low-symmetric martensite phase. Hence, DOS just at the Fermi level decrease in the martensite phase simultaneously increases in regions with higher energy than the Fermi level, such as the energy region of B. The spectral changes at the peaks of C and D are also attributed to the difference in the atomic distances in the martensite and austenite phases, although the reported experimental spectra do not agree with the theoretical ones. The spectra obtained here are consistent with the reported spectra with regard to C and D. Figure 8 (c) shows the $\Delta XAS_{Av}$ in Co-$L_3$ edge. It is difficult to discuss the spectral feature because of the low accuracy, however, a significant change is observed relative to that in the Mn-$L_3$ edge. There is no report yet on the partial DOS of Co in Ni-Co-Mn-In alloys; hence, the role of Co atoms in the martensitic transformation has not been clarified. It is inferred that the DOS would decrease at the Fermi level in the martensite phase, as in the similar trend with the Ni (See the insets of Fig. 5 (a) and 6 (a)).

## IV. Conclusions

In this study, X-ray absorption spectroscopy (XAS) and the magnetic circular dichroism (MCD) for the $Ni_{45}Co_5Mn_{36.7}In_{13.3}$ metamagnetic shape memory alloy were investigated under magnetic fields up to 12 T with a pulsed magnet. Large changes in the MCD signals were observed in all the magnetic elements, Mn, Ni, and Co, due to the field–induced reverse martensitic transformation from the weak-magnetic martensite phase to the ferromagnetic austenite phase. The sign of the MCD signals reveals that Mn, Ni, and Co couple ferromagnetically in the austenite phase. The spin and orbital magnetic momenta, $M_{spin}$ and $M_{orb}$, respectively, of Mn atom in the induced austenite phase, as estimated using the sum rule, were 3.2 and 0.13 $\mu_B$, respectively, resulting in the $M_{orb} / M_{spin}$ ratio of 0.04. In each element, the value of $M_{orb} / M_{spin}$ in the martensite phase with a lower symmetric structure tends to be larger than that in the austenite phase with a cubic structure. The total magnetic moment at 150 K in the induced austenite phase was 6.0 $\mu_B$/f.u., which is comparable to 6.3 $\mu_B$/f.u. obtained from the magnetization curve.

From the element-specific magnetization obtained by plotting the MCD intensities, metamagnetic behavior associated with the field–induced reverse transformation is clearly observed and the reverse transformation finishing magnetic field and martensitic transformation starting magnetic field are detected, although the transformation hysteresis is slightly expanded compare to the magnetization curves measured at the same temperature.

The averaged XAS spectra in parallel and antiparallel configurations, relative to the magnetization direction, were compared in the martensite and austenite phases to clarify the intrinsic spectral change associated with the structural transformation without magnetic contributions to the spectra. There was no visible spectral change in the Mn-$L_3$ edge, whereas it was evident in Ni. This suggests that Ni-3*d* electrons play a more important role in the martensitic transformation than Mn-3*d* electrons. The large spectral change at the peak top of the Ni-$L_3$ edge corresponds to the change in the DOS at the Fermi level, and this is consistent with the decrease in DOS in the martensite phase, as predicted by theoretical calculations. The magnetic field–induced austenite phase in this present study indicates a similar electronic state as the thermal induced austenite phase. Our results show the possibility and effectiveness of soft X-ray XAS-MCD experiments under higher magnetic fields with pulsed magnets.

**ACKNOWLEDGMENTS**

We thank K. Kindo for his technical support in the installation of the pulse magnet. This study was supported by Grant-in-Aids from the Japanese Society for the Promotion of Science (JSPS), Ministry of Education, Culture, Sports, Science and Technology (MEXT), Japan. The synchrotron radiation experiments were conducted at BL25SU of SPring-8 with the approval of the Japan Synchrotron Radiation Research Institute (JASRI) (Proposal No. 2015B1345). Partial fundamental measurements were carried out at the High Field Laboratory for Superconducting Materials and the Center for Low Temperature Science Institute for Materials Research, Tohoku University.

**Author Contributions:** R.Y.U. and R.K. conceived and designed the experiments; R.Y.U. fabricated the specimen and carried out the magnetic measurements; H.Y., Y.N., Y.K., T.N., and H.N. performed the XAS experiments; R.Y.U. analyzed the data and wrote the paper; All the authors participated in the discussions, and the manuscript was finalized under their agreements.

Table Number of 3$d$ electrons, $n_d$, spin magnetic moment, $m_{spin}$, orbital magnetic moment, $m_{orb}$, and $m_{orb} / m_{spin}$ ratio, obtained by applying the sum rule to the MCD spectra in 1 T (martensite phase) and 12 T (austenite phase) at 150 K for $Ni_{45}Co_5Mn_{16.7}In_{13.3}$.

|     | $n_d$ | Magnetic field | $m_{spin}$ ($\mu_B$) | $m_{orb}$ ($\mu_B$) | $m_{orb} / m_{spin}$ |
|-----|-------|----------------|----------------------|---------------------|----------------------|
| Mn  | 5.17  | 1 T            | 1.00                 | 0.07                | 0.08                 |
|     |       | 12 T           | 3.19                 | 0.08                | 0.03                 |
| Ni  | 8.50  | 1 T            | 0.24                 | 0.04                | 0.16                 |
|     |       | 12 T           | 0.47                 | 0.06                | 0.13                 |
| Co  | 7.51  | 1 T            | 0.18                 | 0.02                | 0.11                 |
|     |       | 12 T           | 1.25                 | 0.03                | 0.03                 |

Captions

Fig. 1 Thermomagnetization (*M-T*) curves measured under the magnetic fields of (a) 0.05, 1, 3 T, and (b) of 5 and 9 T, respectively. $T_{Af}$ and $T_{Ms}$ denote the reverse transformation finishing temperature to the austenite phase and the martensitic transformation starting temperature, respectively.

Fig. 2 Magnetization (*M-H*) curves at (a) 100, 150, and 200 K and (b) at 7, 50, and 100 K. $H_{Af}$ and $H_{Ms}$ denote field-induced transformation finishing magnetic field to austenite phase and martensitic phase transformation starting magnetic field, respectively.

Fig. 3 Magnetic field vs. temperature phase diagram for $Ni_{45}Co_5Mn_{36.7}In_{13.3}$. $T_{Af}$ and $T_{Ms}$ are obtained from Fig. 1 (a) and (b), and $H_{Af}$ and $H_{Ms}$ from Fig. 2 (a) and (b). $T_0 = (T_{Af} + T_{Ms})/2$ and $H_0 = (H_{Af} + H_{Ms})/2$.

Fig. 4 XAS and MCD spectra for Mn-$L_{2,3}$ edge with a pulsed magnet in $Ni_{45}Co_5Mn_{36.7}In_{13.3}$ measured at 150 K. $\mu^-$ and $\mu^+$ correspond to the spectra obtained at antiparallel and parallel configuration relative to the magnetization direction, respectively. (a) averaged XAS spectra, $XAS_{Av.}$, recorded under 1-3 T (10-12 T) magnetic fields. Inset is the expanded scale for the $XAS_{Av.}$ peak top of the Mn-$L_3$ edge. The phase states at 1-3 and 10-12 T correspond with martensite and austenite phases, respectively. (b) XAS spectra of $\mu^+$ and $\mu^-$ at 10-12 T, and MCD spectra at 1-3, 4-6, 7-9, and 10-12 T for Mn-$L_{2,3}$ edge.

Fig. 5 XAS and MCD spectra for Ni-$L_{2,3}$ edge with a pulsed magnet in $Ni_{45}Co_5Mn_{36.7}In_{13.3}$ measured at 150 K. (a) averaged XAS spectra, $XAS_{Av.}$, recorded under 1-3 (austenite phase) and 10-12 T (martensite phase) magnetic fields. The notations are the same as in the Fig. 4. Arrows A, B, C, D indicate the same photon energies reported by Klaer *et al*. [23]. Inset is the expanded scale for the $XAS_{Av.}$ peak top of the Ni-$L_3$ edge. (b) XAS spectra of $\mu^+$ and $\mu^-$ at 10-12 T, and MCD spectra in 1-3, 4-6, 7-9, and 10-12 T for Ni-$L_{2,3}$ edge.

Fig. 6   XAS and MCD spectra for Co-$L_{2,3}$ edge with a pulsed magnet in $Ni_{45}Co_5Mn_{36.7}In_{13.3}$ measured at 150 K. (a) averaged spectra, $XAS_{Av.}$, obtained under 1-3 (austenite phase) and 10-12 T (martensite phase) magnetic fields. The notations are the same as in Fig. 4. Inset is the expanded scale for the $XAS_{Av.}$ peak top of the Co-$L_3$ edge. (b) XAS spectra of $\mu^+$ and $\mu^-$ at 10-12 T, and MCD spectra at 1-3, 4-6, 7-9, and 10-12 T for Co-$L_{2,3}$ edge.

Fig. 7   (a) Element-specific hysteresis of Mn, Ni, and Co obtained from the maximum intensities at each $L_3$ edge MCD spectra at 150 K. The photon energy is 639.6, 852.6, and 778.1 eV for Mn, Ni, and Co, respectively. (b) M-H curve obtained at 150 K by magnetic measurement using superconducting magnet. The curve is extracted from Fig. 2 (a). $H_{As}$ and $H_{Af}$ denote reverse transformation starting and finishing magnetic field to the austenite phase, respectively, and $H_{Ms}$ and $H_{Mf}$ the martensitic transformation starting and finishing magnetic field, respectively.

Fig. 8   Difference in the averaged XAS spectra, $\Delta XAS_{Av.}$, between at 10-12 and 1-3 T for Mn, Ni, and Co obtained from Fig. 4(a), 5(a), and 6(a), respectively. The phase states at 10-12 and 1-3 T correspond with austenite and martensite phases, respectively.

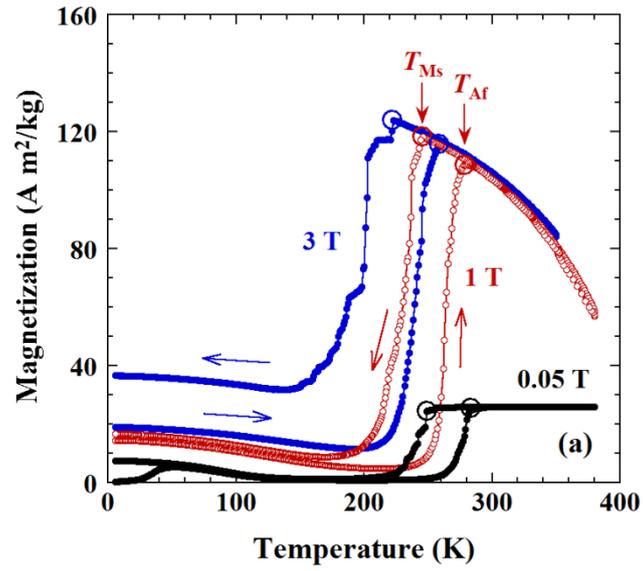

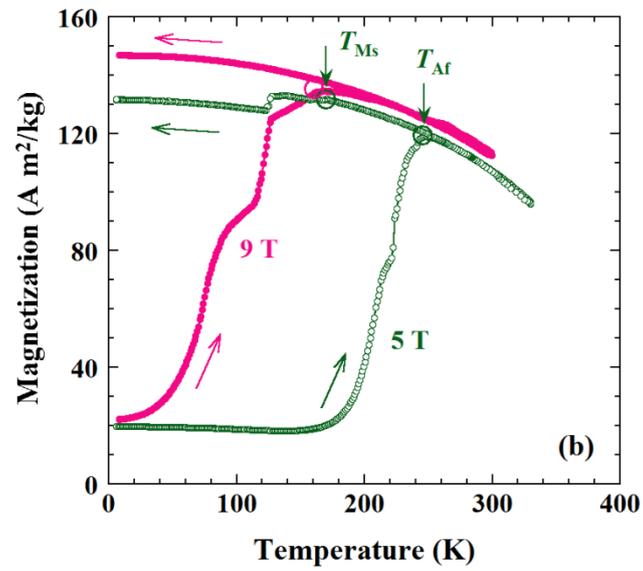

Fig. 1 (a) and (b)

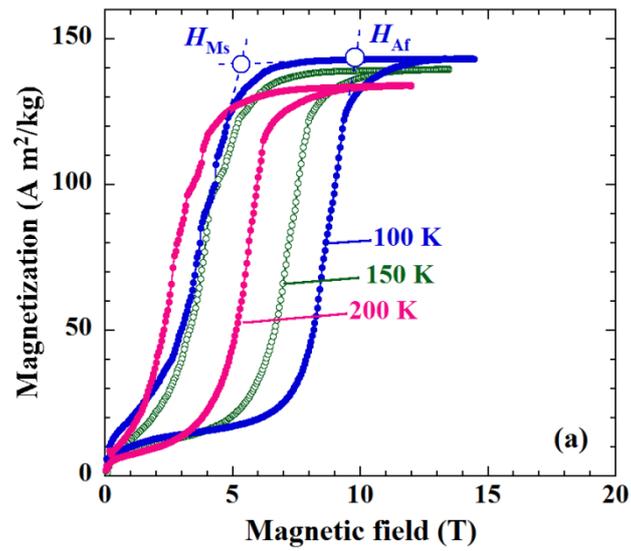

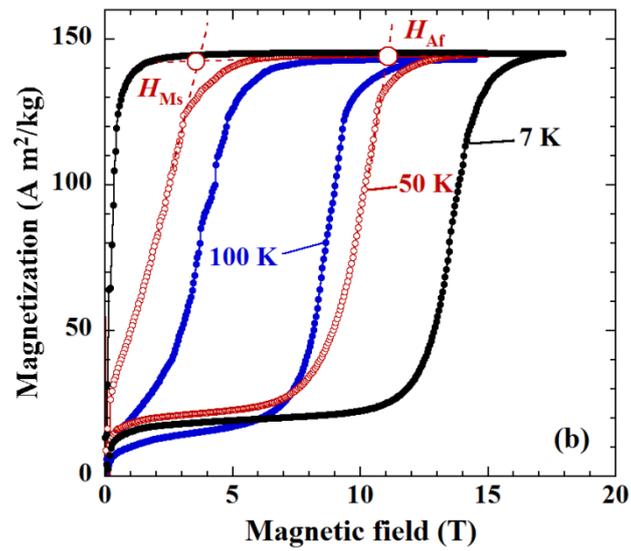

Fig. 2 (a) and (b)

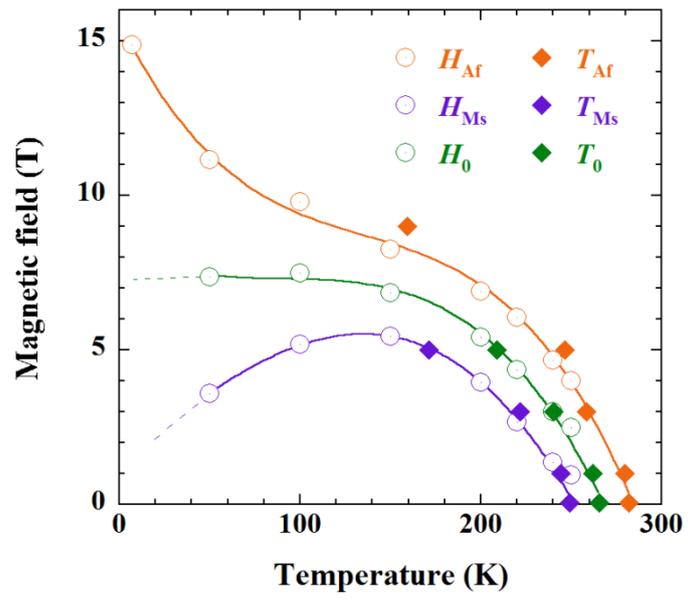

Fig. 3

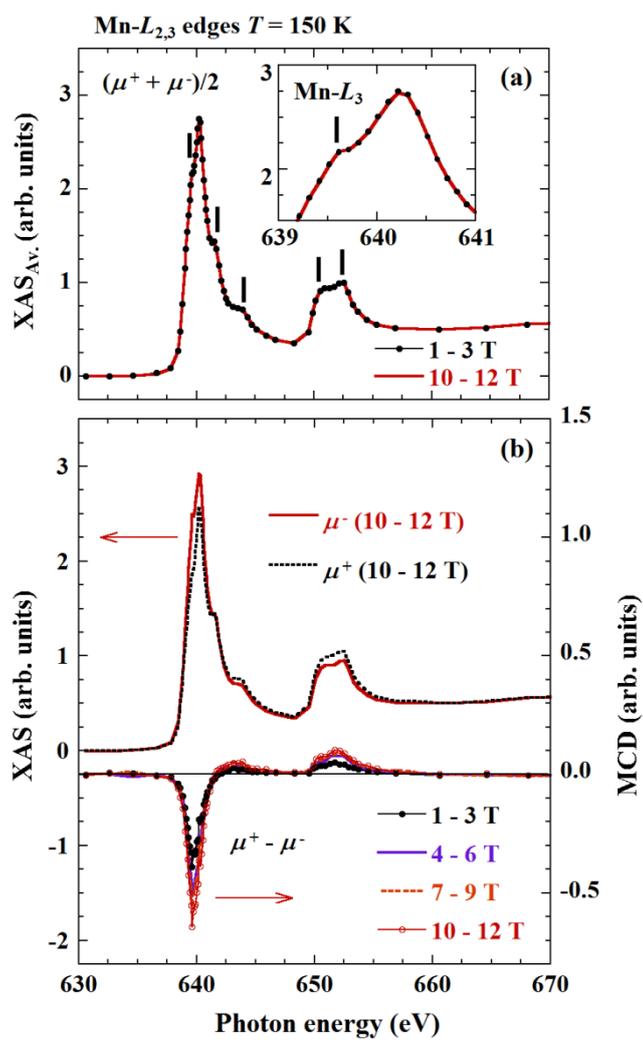

Fig. 4 (a) and (b)

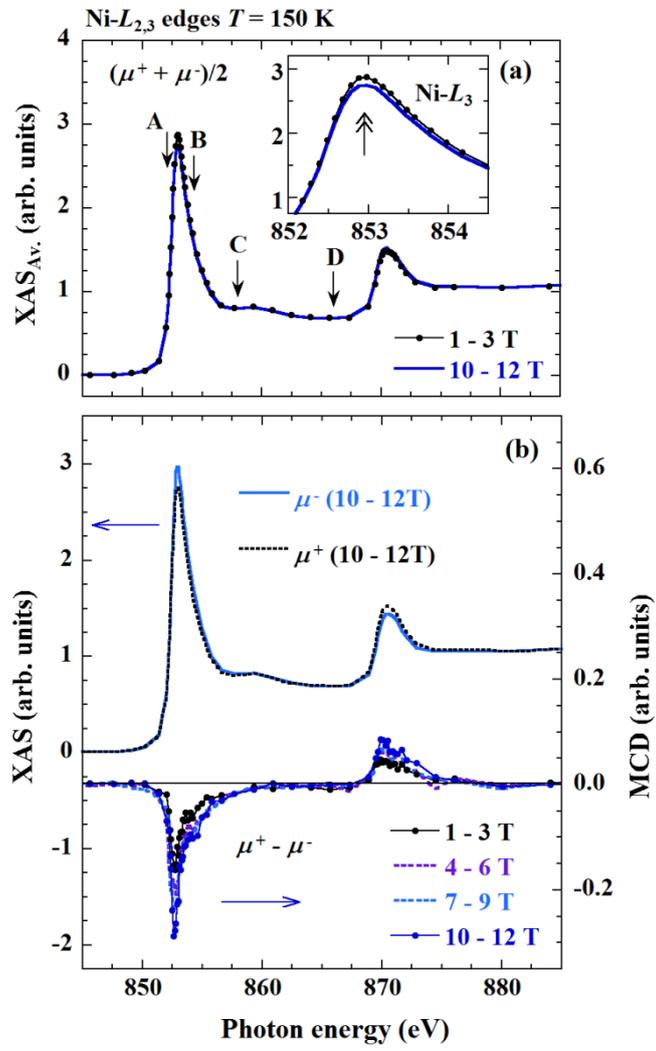

Fig. 5 (a) and (b)

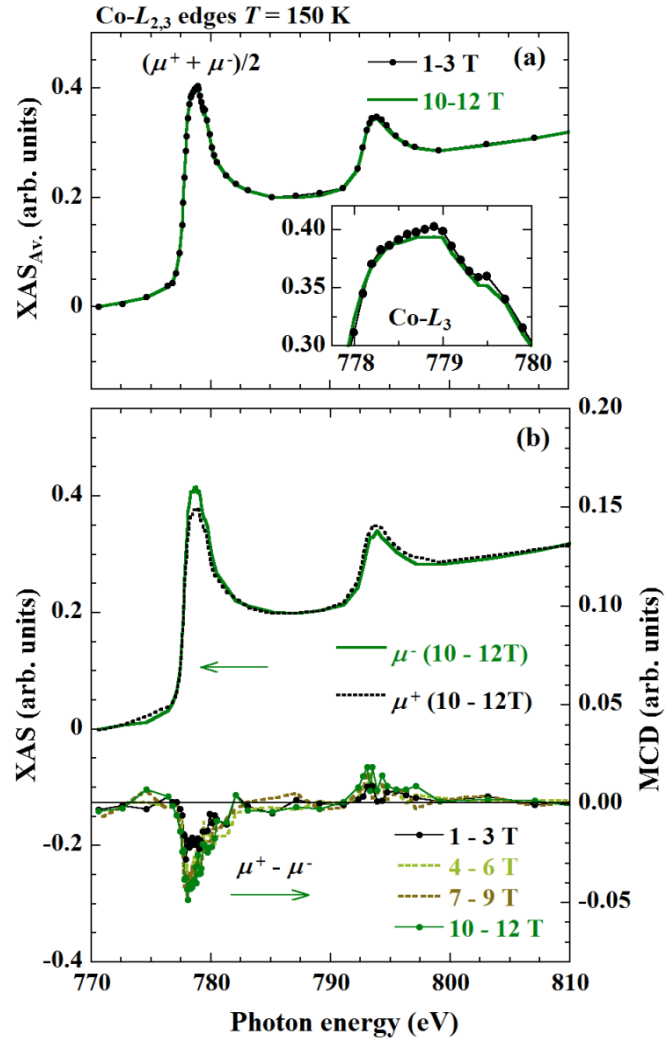

Fig. 6 (a) and (b)

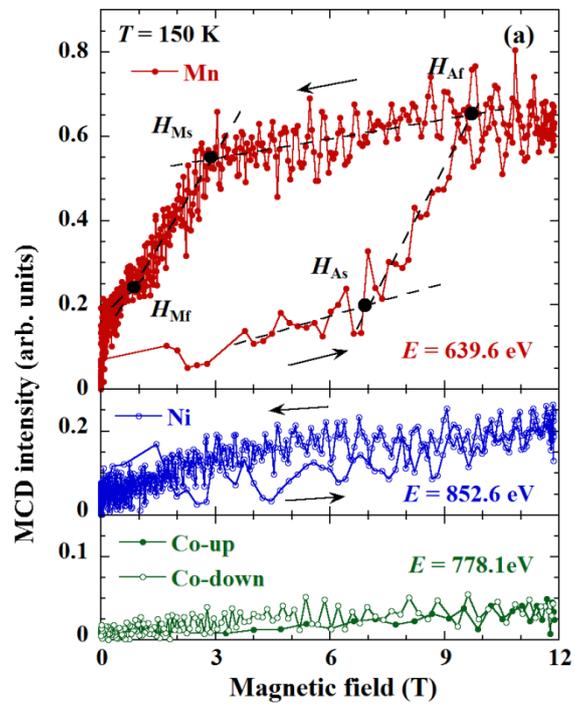

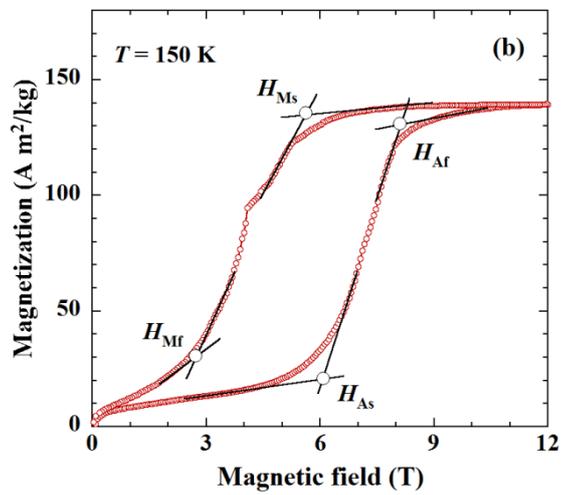

Fig. 7 (a) and (b)

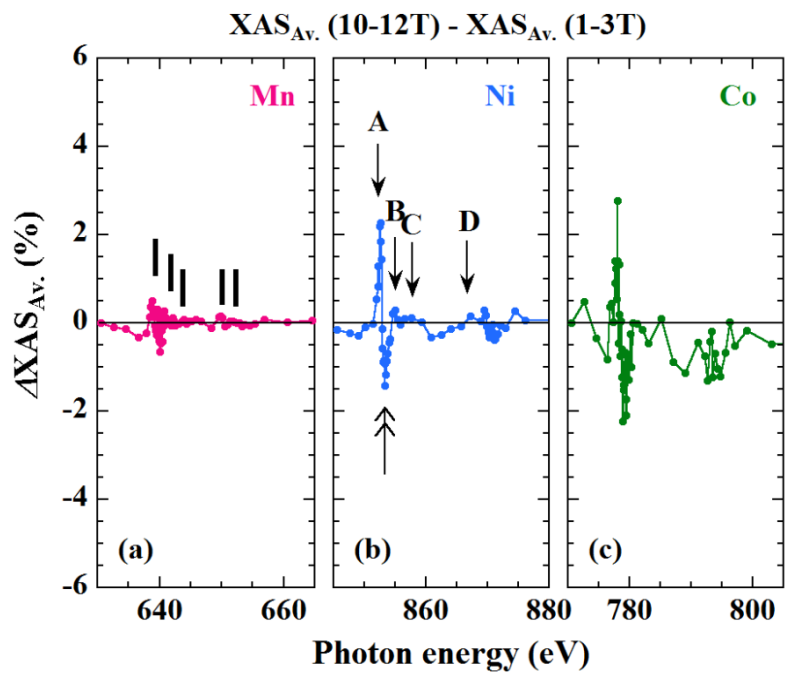

Fig. 8 (a), (b) and (c)